\documentclass[sigconf, manuscript]{acmart}
\AtBeginDocument{%
  }

\setcopyright{acmlicensed}
\copyrightyear{2018}
\acmYear{2018}
\acmDOI{XXXXXXX.XXXXXXX}
\acmConference[Conference acronym 'XX]{Make sure to enter the correct
  conference title from your rights confirmation email}{June 03--05,
  2018}{Woodstock, NY}
\acmISBN{978-1-4503-XXXX-X/2018/06}

\begin{document}

\title{LLM impact on BLV programming}

\author{Prashant Chandrasekar}
\affiliation{%
  \institution{The University of Mary Washington}
  \city{Fredericksburg}
  \state{Virginia}
  \country{USA}}
\email{pchandra@umw.edu}

\author{Mariel Couvillion}
\affiliation{%
  \institution{The University of Mary Washington}
  \city{Fredericksburg}
  \state{Virginia}
  \country{USA}}
\email{mcouvill@mail.umw.edu}

\author{Ayshwarya Saktheeswaran}
\affiliation{%
  \institution{Microsoft Inc.}
  \country{USA}}
\email{ashsa@microsoft.com}

\author{Jessica Zeitz}
\affiliation{%
  \institution{The University of Mary Washington}
  \city{Fredericksburg}
  \state{Virginia}
  \country{USA}}
\email{jzeitz@umw.edu}

\renewcommand{\shortauthors}{Chandrasekar et al.}

\begin{abstract}

Large Language Models (LLMs) are rapidly becoming integral to a wide range of tools, tasks, and problem-solving processes, especially in software development.
Originally designed for natural language processing tasks such as text generation, LLMs are increasingly being used to assist both professionals and students in writing code.
This growing reliance on LLM-based tools is reshaping programming workflows and task execution.
In this study, we explore the impact of these technologies on blind and low-vision (BLV) developers.
Our review of existing literature indicates that while LLMs help mitigate some of the challenges faced by BLV programmers, they also introduce new forms of inaccessibility.
We conducted an evaluation of five popular LLM-powered integrated development environments (IDEs), assessing their performance across a comprehensive set of programming tasks.
Our findings highlight several unsupported scenarios, instances of incorrect model output, and notable limitations in interaction support for specific tasks.
Through observing BLV developers as they engaged in coding activities, we uncovered key interaction barriers that go beyond model accuracy or code generation quality.
This paper outlines the challenges and corresponding opportunities for improving accessibility in the context of generative AI-assisted programming.
Addressing these issues can meaningfully enhance the programming experience for BLV developers. As the generative AI revolution continues to unfold, it must also address the unique burdens faced by this community.

\end{abstract}



\maketitle

\section{Introduction} \label{intro}

As LLMs and GenerativeAI (GenAI) have become more widely used by the public, they can be used by BLV developers to address difficulties when coding (shown in Figure \ref{fig:Solutions and Problems Introduced by LLMs}).
However, LLMs and GenAI introduce new problems for users and often contain accessibility issues \cite{Adnin2024, Seo2024, Cheema2024, Alharbi2024, Pang2025, Penuela2025}.
One significant issue is the difficulty BLV users may face in verifying the correctness of AI-generated code or explanations, as they cannot easily rely on visual inspection or compare results with visual representations \cite{Adnin2024, Penuela2025, Seo2024}. 
In some cases, rather than simplifying tasks, these tools may complicate workflows by requiring users to navigate between multiple GenAI platforms and development environments just to read or edit responses. 
This can ultimately undermine the efficiency gains these technologies aim to provide \cite{Adnin2024}.

To evaluate the real-world impact, we assessed the performance of five popular LLM-powered Integrated Development Environments (IDEs) across a comprehensive set of programming tasks. 
This evaluation revealed various scenarios that were not adequately supported by these IDEs, as well as instances where the LLMs provided incorrect responses. 
A significant finding was the lack of standardization in LLM output across different IDEs for the same prompt, forcing BLV developers to grapple with the challenge of skimming these diverse responses.

We conducted a user study to observe the first-hand experiences of BLV individuals using LLMs for coding-related tasks. 
The study involved participants working with two Python programs with introduced errors while their programming behavior, tool usage, and ChatGPT queries were observed. 
The goal was to understand the integration of LLMs into their existing workflows as they tackled code comprehension, editing, debugging, skimming, and navigation tasks. 
This direct engagement revealed specific issues such as difficulties in testing program behavior, unexpected reliance on ChatGPT's potentially incorrect suggestions, and an apparent tendency to overlook detailed code comments, all contributing to a deeper understanding of the obstacles faced by BLV developers.

The remainder of this paper is structured as follows: Section \ref{lit} provides a review of relevant literature. 
Section \ref{sec:second-hand-eval} details the results from our evaluation of LLM-powered IDEs. 
Section \ref{sec:user-study} details the study setup, including participant recruitment and the tasks assigned. 
We highlight key insights derived observations of the BLV developers' programming behaviors. 
We collates the learnings from these interactions, alongside an overview of the current integration of LLMs into development workflows, to offer suggestions for future efforts aimed at better supporting the BLV community in programming. 
It is critical that the generative AI ``revolution'' must address the specific burdens faced by BLV developers to ensure equitable access to these powerful tools.

\section{Related Efforts} \label{lit}

\begin{table}
    \centering
    \caption{Solutions and Problems Introduced by LLMs}
    \label{fig:Solutions and Problems Introduced by LLMs}
    \begin{tabular}{|p{0.015\linewidth} | p{0.6\linewidth}| p{0.15\linewidth} | p{0.080\linewidth}| p{0.15\linewidth}|}
        \hline
        ID &  Issues &  References&  LLM-fix?& LLM References\\
        \hline
        1 & GUI interfaces (featutes/buttons) in IDEs cannot be easily navigated by BLV users & \cite{Aboubakar2022, Potluri2018} & Yes & \cite{Kodandaram2024} \\
        \hline
        2 & Table outputs from CLIs are not easily interpreted by screen readers & \cite{Sampath2021} & Yes & \cite{Alharbi2024} \\
        \hline
        3 & Screen readers make it difficult to navigate, skim, comprehend, edit, or debug code & \cite{Aboubakar2022, Albusays2016, Mealin2012} & Yes & \cite{Kodandaram2024, Seo2024} \\
        \hline
        4 & BLV developers cannot easily skim code & \cite{Sampath2021, Aboubakar2022, Albusays2016} & Yes & \cite{Adnin2024, Pang2025} \\
        \hline
        5 & Finding a specific element in the code can be difficult; once found, developers may lose their original place in the code & \cite{Mealin2012} & Yes & \cite{Adnin2024, Pang2025} \\
        \hline
        6 & Attempting to understand the structural information about code is hard to do with a screen reader & \cite{Baker2015} & Yes & \cite{Adnin2024, Pang2025} \\
        \hline
        7 & Screen readers make it difficult to gain a holistic view of the code base & \cite{Albusays2016, Baker2015} & Yes & \cite{Adnin2024, Pang2025} \\
        \hline
        8 & Screen readers make it difficult to comprehend what a program is doing & \cite{Stefik2007} & Yes & \cite{Adnin2024, Pang2025} \\
        \hline
        9 & Most coding tools created for BLV developers often rely on providing spatial context when semantic context is more valuable & \cite{Andreas2009} & Yes & \cite{Adnin2024, Pang2025} \\
        \hline
        10 & Most built-in debugging tools are inaccessible, so many BLV developers rely on basic techniques & \cite{Khaled2017, Mealin2012, Albusays2016, Sampath2021, Aboubakar2022, Potluri2018, Baker2015} & Yes & \cite{Adnin2024, Pang2025}  \\
        \hline
        11 & Modern IDEs have inaccessible debugging tools & \cite{Aboubakar2022, Albusays2016} & Yes & \cite{Kodandaram2024} \\
        \hline
        12 & BLV developers often lose their place when editing code directly in the source file & \cite{Aboubakar2022} & Yes & \cite{Adnin2024, Pang2025} \\
        \hline
        13 & Many BLV developers will copy a section of code into a temporary buffer, make edits there, and paste the modified code back into the file & \cite{Sampath2021, Mealin2012, Aboubakar2022} & No &  \\
        \hline
        14 & BLV developers may accidentally add changes to code in unintended locations & \cite{Aboubakar2022} & No & N/A \\
        \hline
        15 & Hallucinations in terms of images analysis & \cite{Adnin2024, Avestimehr2025, Seo2024, Avestimehr2025} & N/A & N/A \\
        \hline
        16 & Hallucinations in text analyses and text generation & \cite{Adnin2024, Penuela2025, Pang2025} & N/A & N/A \\
        \hline
        17 & LLMs and AI can be inaccessible to BLV users & \cite{Adnin2024, Seo2024, Cheema2024, Alharbi2024, Pang2025, Penuela2025} & N/A & N/A \\
        \hline
    \end{tabular}
    
\end{table}

\subsection{Can LLMs Provide Solutions?}
Many issues that BLV developers face when using screen readers, debugging, or editing code can be alleviated—if not entirely solved—by leveraging the basic capabilities of GenAI. 
Screen readers often make it difficult to navigate, skim, comprehend, edit, and debug code effectively \cite{Aboubakar2022, Albusays2016, Mealin2012}. 
When encountering challenges in completing coding tasks, AI and LLMs can assist BLV developers by providing code summaries, identifying specific line numbers, flagging errors, and offering targeted support \cite{Adnin2024, Kodandaram2024, Seo2024}. However, due to the inherently linear nature of screen readers, these tools cannot fully eliminate the underlying accessibility barriers.
Debugging, in particular, remains a frustrating task for BLV developers. 
Built-in debugging tools are often inaccessible, leading many to rely on basic techniques like inserting print statements to trace program behavior \cite{Khaled2017, Mealin2012, Albusays2016, Sampath2021, Aboubakar2022, Potluri2018, Baker2015}. 
GenAI can help reduce this burden by identifying and fixing errors or bugs in the code, decreasing dependence on manual debugging methods \cite{Adnin2024, Pang2025}. 
This support can also be valuable during code editing, where modifying a line in the source file can cause developers to lose their place or context \cite{Aboubakar2022}. 
By prompting GenAI to make changes or locate errors, developers can stay oriented within the file and maintain their workflow more easily \cite{Adnin2024, Pang2025}.

Another way that GenAI and LLMs can support BLV developers is by generating overviews and summaries of code.
These are particularly useful for helping BLV developers navigate, skim, or understand a program’s overall purpose and structure.
These are tasks that are otherwise difficult due to the limitations of screen readers. 
Skimming through code, for instance, can be nearly impossible without visual cues \cite{Sampath2021, Aboubakar2022, Albusays2016}. 
Understanding structural elements such as function definitions or loops also poses a challenge when auditory tools are the only interface \cite{Baker2015}. 
Additionally, screen readers make it hard to develop a holistic mental model of a codebase \cite{Albusays2016, Baker2015} and to grasp a program’s high-level behavior \cite{Stefik2007}. 
In these scenarios, GenAI can offer valuable assistance by summarizing code structure, describing program functionality, or locating specific elements within the file.
Thus, ultimately making codebases more accessible to BLV developers \cite{Adnin2024, Pang2025}.

AI and LLMs can assist BLV developers by translating visually presented or spatially organized information into descriptive, semantic content that is more accessible through screen readers.
For example, CLI outputs, especially those formatted as tables, are often inaccessible and difficult for screen readers to interpret \cite{Sampath2021}. 
GenAI tools, such as AI VAT, can help by generating textual descriptions of these tables, making CLI outputs easier to understand \cite{Alharbi2024}.
However, these tools tend to struggle with recognizing and interpreting information presented in table formats which leads to disorganized and unusable output \cite{Alharbi2024}.
Another common challenge for BLV developers is locating specific elements in code, such as variable names or definitions. 
Once an element is found, it can be easy to lose track of the original position in the file \cite{Mealin2012}. 
GenAI can assist by responding to prompts to locate, describe, or summarize code elements
which in turn makes it easier to explore and navigate code without losing context \cite{Adnin2024, Pang2025}.
Additionally, many tools designed for BLV developers focus on providing spatial context, even though semantic context is often more useful for understanding and working with code \cite{Andreas2009}. 
AI tools can be prompted to generate semantic descriptions of code, helping developers understand its meaning and functionality \cite{Adnin2024, Pang2025}.

Some AI-based tools are specifically designed to help BLV individuals interpret visual data and navigate complex digital environments, including those commonly found in software development.
GUI interfaces in IDEs, for example, are often inaccessible to BLV users \cite{Aboubakar2022, Potluri2018}. 
In addition to interface navigation, modern IDEs frequently include debugging tools that are not designed with accessibility in mind \cite{Aboubakar2022, Albusays2016}.
However, with the help of AI and LLM-based tools like Savant \cite{Kodandaram2024}, users can interact with application interfaces using natural language commands. 
This capability allows BLV developers to more easily and accessibly navigate and operate within GUI-based environments.

While AI and LLMs can assist with many challenges faced by BLV developers, they cannot solve every problem.
Some problems stem from broader workflow and interface limitations that AI cannot fully resolve. 
For instance, many BLV developers adopt a workaround where they copy a section of code into a temporary buffer, make edits there, and then paste the modified version back into the original file. 
Although this method allows for more focused editing, it leads to an inefficient and fragmented workflow \cite{Sampath2021, Mealin2012, Aboubakar2022}. 
Even with the assistance of AI for tasks like debugging, this core issue of disjointed interaction with the code remains unresolved.
Another common challenge is the risk of unintentionally editing or inserting code in the wrong location, which can happen easily when relying on screen readers alone \cite{Aboubakar2022}. 
While AI might help identify resulting errors, such as syntax or logical bugs, it cannot prevent the initial misplacement or streamline the editing process in a way that fully eliminates this risk.

\subsection{Problems Introduced with LLMs}
A major concern is the risk of hallucinations, or when a GenAI tool provides inaccurate or fabricated information \cite{Adnin2024, Avestimehr2025, Seo2024, Penuela2025, Pang2025}.
For BLV developers, the impact of such hallucinations can be especially disruptive. 
Unlike sighted users who can quickly scan and validate outputs, BLV users may struggle to verify the accuracy of results, further complicating the development process \cite{Seo2024, Cheema2024, Adnin2024, Alharbi2024}. 
Even when responses are purely textual, validation often requires additional steps like switching applications or conducting follow-up searches.
These are tasks that can be tedious and difficult when compounded by accessibility barriers in navigating and copying content across platforms \cite{Pang2025, Adnin2024, Alharbi2024}.

Moreover, many LLM and AI-based tools themselves are not fully accessible to BLV users, which can limit or negate their usefulness altogether \cite{Adnin2024, Seo2024, Cheema2024, Alharbi2024, Pang2025, Penuela2025}.
Common accessibility barriers—such as unlabeled buttons, poor screen reader support, and difficult navigation—make using these tools frustrating or even impossible for some BLV individuals \cite{Adnin2024}. 
Although some platforms offer partial accommodations, they are often not designed with BLV users in mind, resulting in cumbersome and inefficient user experiences. 
As reliance on GenAI continues to grow, it is crucial to address these verification and accessibility challenges to ensure that these tools benefit all developers equitably, including those in the BLV community.

\section{Second-Hand Evaluation of LLM-powered tools} \label{sec:second-hand-eval}

In this section, we describe our efforts to measure LLM support for programming tasks (such as code comprehension, code editing, code debugging, among others) and sub-tasks.
We do this in order to understand the scope and extent of support of LLMs and, more interestingly, scenarios where support is lacking and/or scenarios were models are inconsistent in their correctness. 

\begin{table}
    \centering
    \caption{(Sub-)Set of Scenarios for LLMs}
    \label{tab: scenariosForLLMs}
    \begin{tabular}{| p{0.20\linewidth}| p{0.80\linewidth}|}
        \hline
        Sub-task &  User Prompt (User Perspective)\\
        \hline
Understanding function purpose and behavior&- I need to modify the find\textunderscore target function so that it removes the target from the data and returns it.
- If the target isn’t found, it should return None.\\
\hline
Correcting indentation errors&- I’m not sure what the correct indentation level is for line 15. Can you help me fix it?\\
\hline
Interpreting class diagrams and inheritance without visual aids&- I’m trying to understand the relationship between the HealthCommand and Command classes.
- Can you explain their inheritance structure in a way that works with my screen reader?\\
\hline
Understanding standard library functions and error-related output&- I want to know what sys.argv represents in this code. Why does the program use the second element instead of the first?
- I’m trying to understand an error message. What does the $os.path.join()$ function do in this context, and how does it relate to the issue?\\
\hline
Interpreting and summarizing long error messages&"- I’ve encountered a Java stack trace. Can you help me identify the function and line that caused the ArithmeticException?
\\
\hline
Understanding loop iteration and behavior&- I’m trying to understand how many times a loop will run in the current code.
- My loop isn't printing six numbers as I expected. Why is that happening?
- I want to know the value of i when p is printed. Also, what gets printed when i equals 18?\\
\hline
Identifying and explaining syntax or logic errors&- I need help identifying what’s wrong with this code.
- Can you explain the cause of the syntax error and where it occurs?\\
\hline
Debugging null pointer or segmentation fault errors&- I'm getting a null pointer error. What happens to the variable that causes this segmentation fault?\\
\hline
Locating and fixing multiple syntax issues&- I want to find and fix all syntax errors in this code.
- Can you guide me through resolving them one by one?\\
\hline

Refactoring methods and updating all references&- I’ve removed a parameter from a method. How can I find and update all the places where this method is called?
- In the count\textunderscore patterns() function, I want to make the limit variable a parameter instead of a local variable. How do I refactor this?\\
\hline
Identifying conditions for deeply nested loop execution&- I need to locate the deepest nested for loop in the code and understand the condition under which it runs.\\
\hline
Locating variable declarations and fixing logic errors&- I want to find the lines where the variables kk, oo, and ee are declared.
- I’m on line 30 and want to know where the variable n (used on line 4) is declared. Can you highlight it for me?\\
\hline
Navigating and customizing IDE interface&- I’m trying to find the “Run” button or open the terminal in my IDE (e.g. VS Code). Where should I look?
- I want each function and class definition to be surrounded by two blank lines instead of one. Can you help me make that formatting change?\\
\hline
Skipping and locating comment blocks&- I want to locate the longest comment in the file, based on line length. Can you find its start and end?\\
\hline
Switching focus between editor and shell&- I want to move my cursor from the code editor to the shell or console window. How can I do that efficiently?\\
\hline
Tracing variable declarations and values&- From line 30, I want to jump to where the variable a is declared (on line 6) and print its value. Can you help?\\
\hline
    \end{tabular}
    
\end{table}

Table \ref{tab: scenariosForLLMs} provides the sub-list of the tasks we asked LLMs to perform in specific coding scenarios.
These scenarios are specifically identified in literature as tasks given to test/measure performance of solutions meant to support BLV developers \cite{Aboubakar2022}.    
We tested LLMs integrated into popular Python IDEs: VSCode, PyCharm, Google Colab, Neovim. We also tested using ChatGPT as an editor.

\begin{table}
    \centering
    \caption{Statistics on support and correctness of LLMs-integrated tools; broken down by programming tasks}
    \label{tab:LLM_Tools_Percent}
    \begin{tabular}{|c|c|c|c|c|}
       \hline
        \textbf{LLM Tool/ Coding Task Type (\# of scenarios)} & \textbf{Comprehension (9)} & \textbf{Debugging (13)} & \textbf{Editing (7)} & \textbf{Navigation (18)}\\
        \hline
        Neovim support & 8/9 & 12/13 & 1/7 & 6/18\\
         \hline
        Neovim inccorect & 1/8 & 1/12 & 1/6 & 3/12\\
         \hline
        VSCode support & 8/9 & 13/13 & 7/7 & 13/18\\
         \hline
        VSCode incorrect & 1/8 & 1/13 & 1/7 & 1/13\\
         \hline
        Google Colab support & 9/9 & 13/13 & 7/7 & 14/18\\
         \hline
        Google Colab incorrect & 0 & 0 & 0 & 2/14\\
         \hline
        PyCharm support & 7/9 & 12/13 & 7/7 & 15/18\\
         \hline
        PyCharm incorrect & 0 & 1/12 & 2/7 & 3/15\\
         \hline
        ChatGPT support & 9/9 & 13/13 & 7/7 & 15/18\\
         \hline
        ChatGPT incorrect & 0 & 0 & 0 & 3/15\\
         \hline
    \end{tabular}

\end{table}

Table \ref{tab:LLM_Tools_Percent} provides the overall numbers for all the LLM-powered tools.
One way to look at this is that the current SOTA LLMs did for the most part (23 out of 31 scenarios) understand and correctly help the users.

\begin{figure}[!htp]
  \centering
  \includegraphics[width=0.75\linewidth]{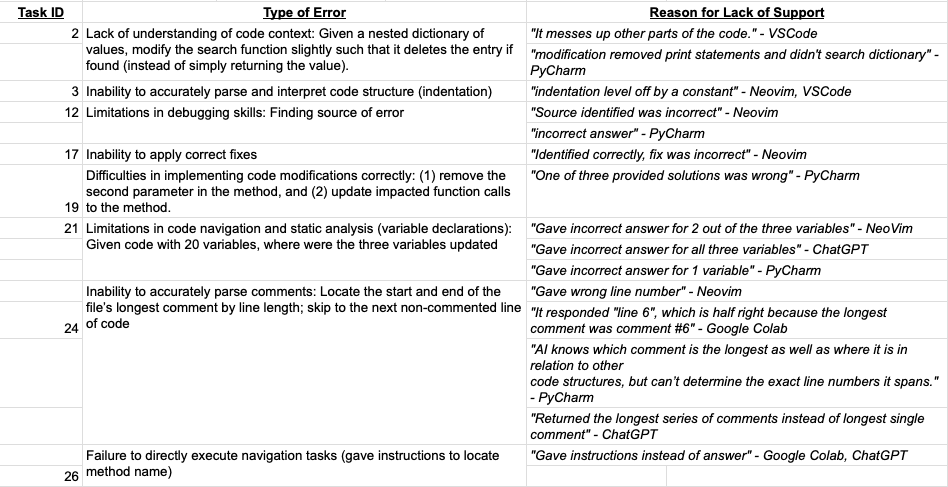}
  \caption{LLMs-integrated tool reasons for lack of support}
  \Description{}
  \label{tab:LLM_ReasonsLackOfSupport}
\end{figure}

Table \ref{tab:LLM_ReasonsLackOfSupport} provide the type of scenarios where the LLMs in the IDE gave inconsistent or incorrect answers.
To be certain, we tried different prompts and simulated these scenarios many times.
These continued to be the most error prone scenarios.

\subsection{``Key'' Observations}

\begin{figure}[!htp]
  \centering
  \includegraphics[width=0.90\linewidth]{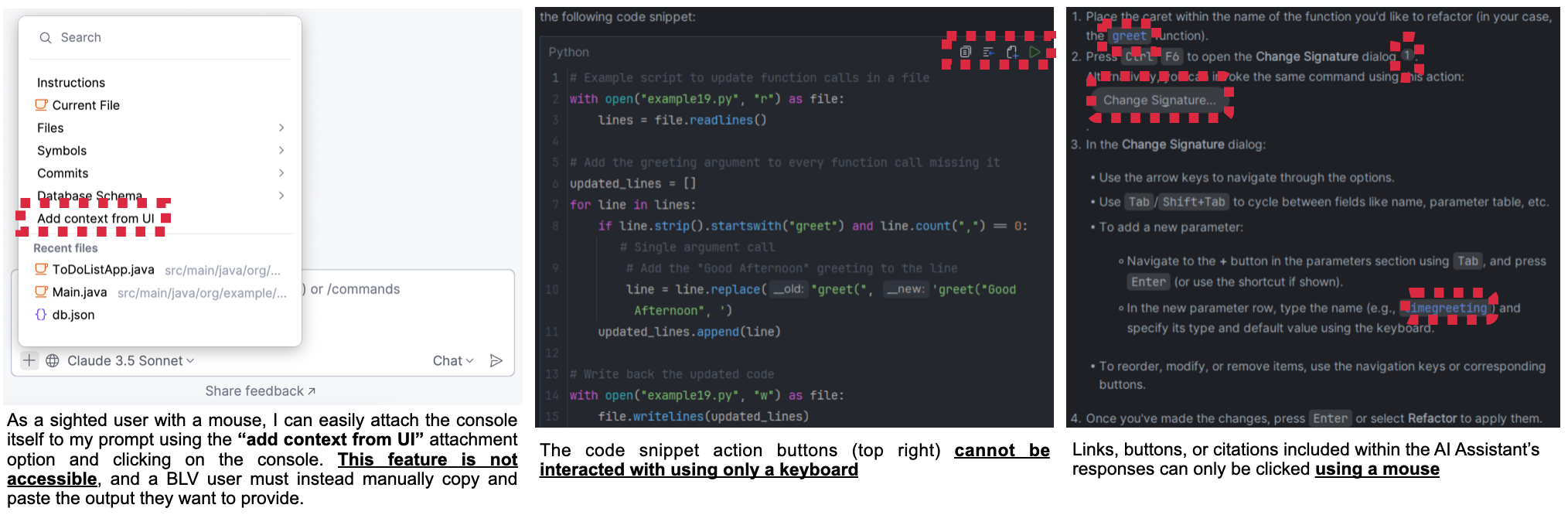}
  \caption{Examples of inaccessibility of LLM response; details provided as well.}
  \Description{}
  \label{fig:LLM_Issues}
\end{figure}

\begin{figure}[!htp]
  \centering
  \includegraphics[width=0.90\linewidth]{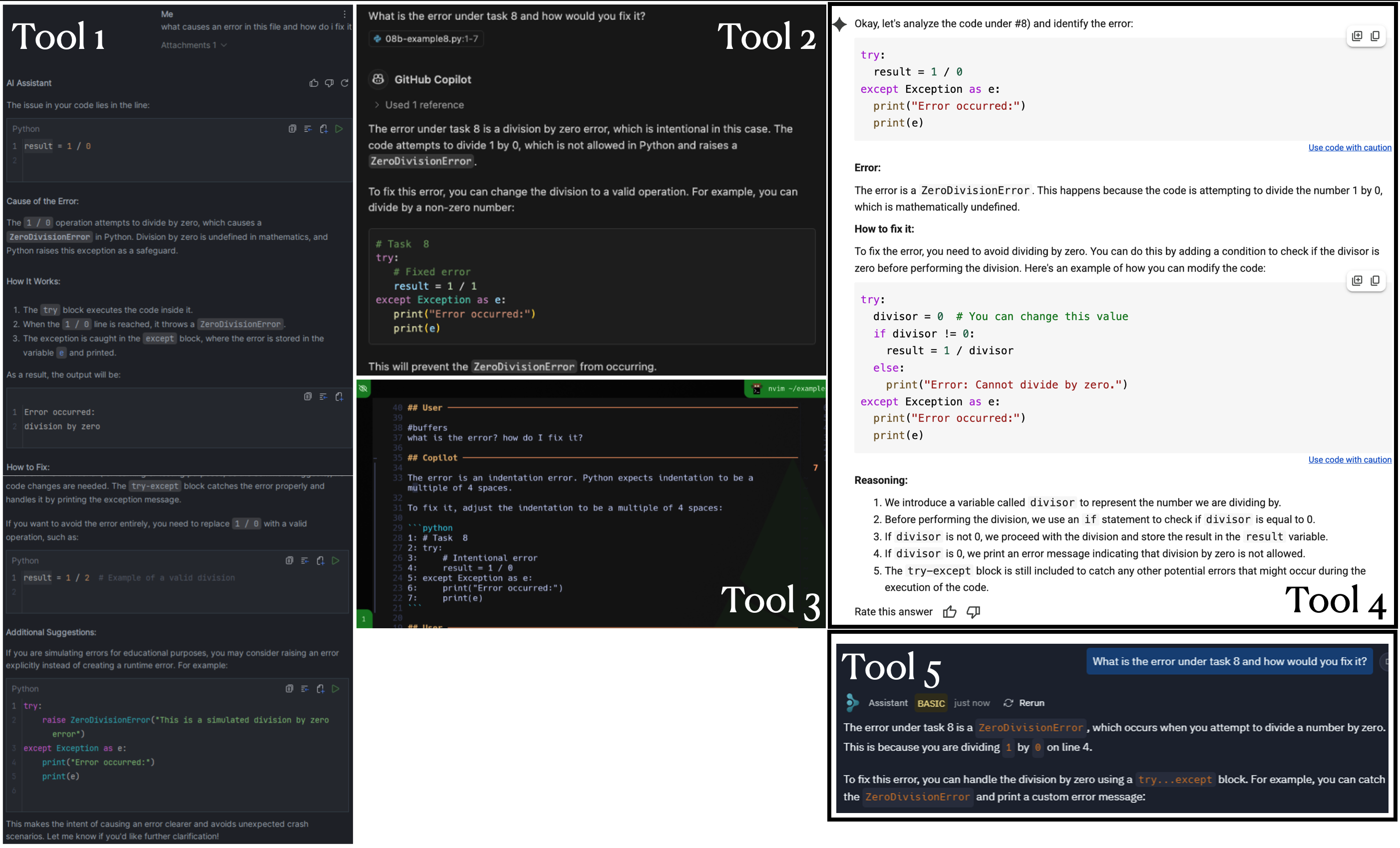}
  \caption{Different LLM-enabled IDE's response to the same prompt}
  \Description{}
  \label{fig:LLM_Response}
\end{figure}

While hallucinations impact both sighted and BLV individuals, verifying the response of LLMs is relatively easier for sighted individuals. 
While models will get better and might better support these scenarios in future versions, no model is 100\% perfect.
Regardless of the model performance, what needs to be addressed is the way the responses are outputted. The output should allow for easy undo-ing/verification/checking.
Specifically, Figure \ref{fig:LLM_Issues} show three different cases where the language of LLM response assumed spatial aware-ness; therefore inaccessible to BLV individuals.
Figure \ref{fig:LLM_Response} showcases the lack of standard or structure in LLM response for the same prompt across the different tools. 
Non-standard response format making skimming/searching challenging; similar to BLV's current experience with unstructured outputs in command-line interfaces. 
Without that core design consideration, nothing changes.

\section{Problems identified by BLV people while completing tasks.} \label{sec:user-study}
The previous sections focused on second-hand insights. We also wanted to witness first-hand experiences of LLM-use for coding-related tasks.

\subsection{User Study}
The goal of the study was to observe (with little-to-no intervention/interruption) LLM-use (as needed) while BLV participants perform code comprehension, code editing, code debugging, code skimming, and code navigation.
We provided two (un-related) Python programs: (1) a solution to a variation of the popular game ``Wordle'' \cite{Wordle2021}, and (2) a solution to a medium-level programming puzzle. 
We selected these programs for specific reasons. First, the Wordle-like project is a common programming project given to students in CS1 and ``Wordle'' is a popular game that requires little code to implement. We hoped that the popularity of the game would increase the participants' familiarity with the game. In addition, the programming puzzle uses a recursive approach that is non-trivial to comprehend and debug.
The goal of the user study was NOT to test the participants' coding ability.
We simply wanted to observe the introduction of LLM-use in their existing workflow as they attempt to answer questions starting with a simple project and moving on to a relatively complicated solution afterwards.

Each Python code was appropriately complemented with comments and user-friendly variable names.
We introduced seven syntactic errors and four logical/functional errors in the first file and five of each in the second file.

The task began with two clear questions about each Python program:
\begin{itemize}
    \item Q1: What problem does the code solve?
    \item Q2: How does the code solve the problem?
\end{itemize}

We also provided the following two prompts:
\begin{itemize}
    \item Prompt 1: If you find any issues in the code, feel free to make changes to address them, so as to help answer the two questions.
    \item Prompt 2: Feel free to ask ChatGPT for help.
\end{itemize}

We did not time their effort for each program.
We wanted to replicate their existing coding environment as much as possible.
Having said that, we ``cut off'' the study after two hours considering fatigue and other factors.

\begin{table}
    \centering
    \caption{Study Details}
    \label{tab:ASSETS_Study}
    \begin{tabular}{|l|l|}
        \hline
        \textbf{Category} & \textbf{Details}\\
         \hline
        BLV Participant Tools  & Preferred IDE, ChatGPT (browser), IDE console, Shell/Terminal, Provided Python file\\
         \hline
        Observer Tools (Us) & Screen sharing, Audio sharing, Screen reader audio sharing\\
         \hline
        What Are We Looking At? & Programming behavior, Tool usage patterns, ChatGPT queries\\
         \hline
        What Are We Measuring? & Log sequential analysis of programming steps, Yule’s Q (behavior patterns), Depth of ChatGPT queries, \\
        & Syntactic error fixes, Logical error fixes, Testing behavior, Responses to post-task questions\\
         \hline
    \end{tabular}
    
\end{table}

Each participant was given the Wordle-like program first. 
When the participant felt they had done enough to address the prompts and have answers to the questions, we stopped, discussed what they completed, and then moved on to the second program to do the same. 
The study ``apparatus'' and measurement details can be found in Table \ref{tab:ASSETS_Study}

\subsection{Participant Details}

We recruited participants for this study with the help of the National Federation of Blind.
They completed two surveys prior to our meeting; the first explaining the motivation, asking consent, and availability, and the second requesting background information about their programming experience, experience with tools, and ChatGPT-like tools for coding tasks.
Five participants completed the study.

\begin{table}
    \centering
    \caption{Participant Details}
    \label{tab:ASSETS_Participants}
    \begin{tabular}{|l|l|l|l|l|}
        \hline
        \textbf{ID} & \textbf{Program for Work or Hobby} & \textbf{Experience} & \textbf{Langauges} & \textbf{Editor}\\
        \hline
        P3 & Hobby & < 5 years & Python,HTML/CSS &  VS Code\\
        \hline
        P4 & Hobby & < 5 years & Python,HTML/CSS,Javascript & Textedit, Notepad, VSCode, \\
        & & & & TexShop, Google Apps Script Editor \\
        \hline
        P5 & Both & 10-20 years & Python,Java,C++,HTML/CSS, & \\
        & & & Javascript, Other &  Notepad++\\
        \hline
        P6 & Job & 5-10 years & Python,HTML/CSS &  VS Code, Curser\\
        \hline
        P7 & Hobby & < 5 years & Python,Javascript &  VS Code\\
        \hline
    \end{tabular}
    
\end{table}

\subsection{``Key'' Observations}

In this paper, we highlight observations that surprised us (pleasantly and unpleasantly).  
More specifically, our observations represent opportunities that would, in our estimation, have the most positive impact to BLV developers in their programming workflow that involves LLMs.

\textbf{Observation 1: Code comprehension deficiency}

One of the fundamental skills we teach in CS1 is the skill to describe the input (I), processing (P), and output (O) behaviors of a program. 
It could be used to describe an existing piece of code, or a framework to help design, build, and test code from ground up. 
Before writing any code, we ask students to build an ``IPO'' chart.
The reason we have them write input and output along with describing the algorithm is to explain ``dynamic'' program behavior for different inputs. 

\begin{figure}[!htp]
  \centering
  \includegraphics[width=0.90\linewidth]{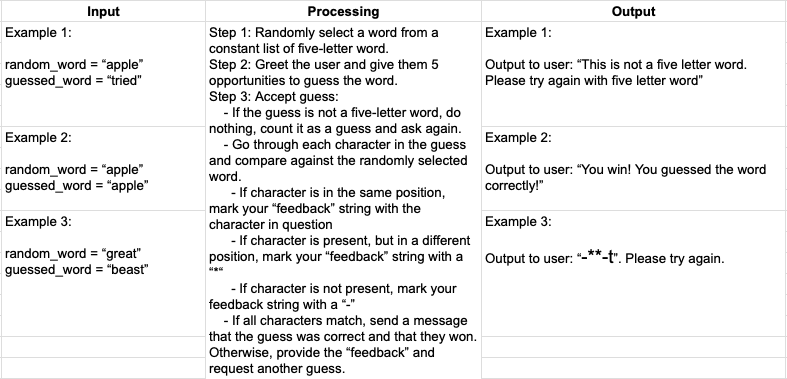}
  \caption{IPO Chart for the first program given to the participants}
  \Description{}
  \label{fig:ASSETS_IPO}
\end{figure}

When error-free, the Wordle-like program works as described in the IPO chart, shown in Figure \ref{fig:ASSETS_IPO}.

All but one participant (P5) \textbf{did not} test the program to see if it gave the correct feedback to their guesses (analogous to Example 3 from \ref{fig:ASSETS_IPO}).
Some tested for incorrect number of letters, some for output after guessing incorrectly five times, and some randomly guessed it correctly. 
However, we were very surprised to see that none of them tested for or verified the correctness of the feedback. 
Participant P5 tried to play the game and guess the word based on the feedback to see if the program worked as intended.

Other literary works have highlighted the challenges BLV users face when using debugging tools within an IDE given the visual nature of the interactive elements (buttons) and the presentation of variable states and program output simultaneously through multiple windows. 
Using a debugger also helps comprehend program behavior.
We hypothesized LLM-integrated IDEs and/or ChatGPT would help ``train'' its users to think about dynamic program behavior.
We uploaded the same program to ChatGPT with the prompt: \textit{``In this Python file is a Wordle-like game. I am a blind programmer. Could you tell me how the program behaves for different sample inputs?''}

\begin{figure}[!htp]
  \centering
  \includegraphics[width=0.50\linewidth]{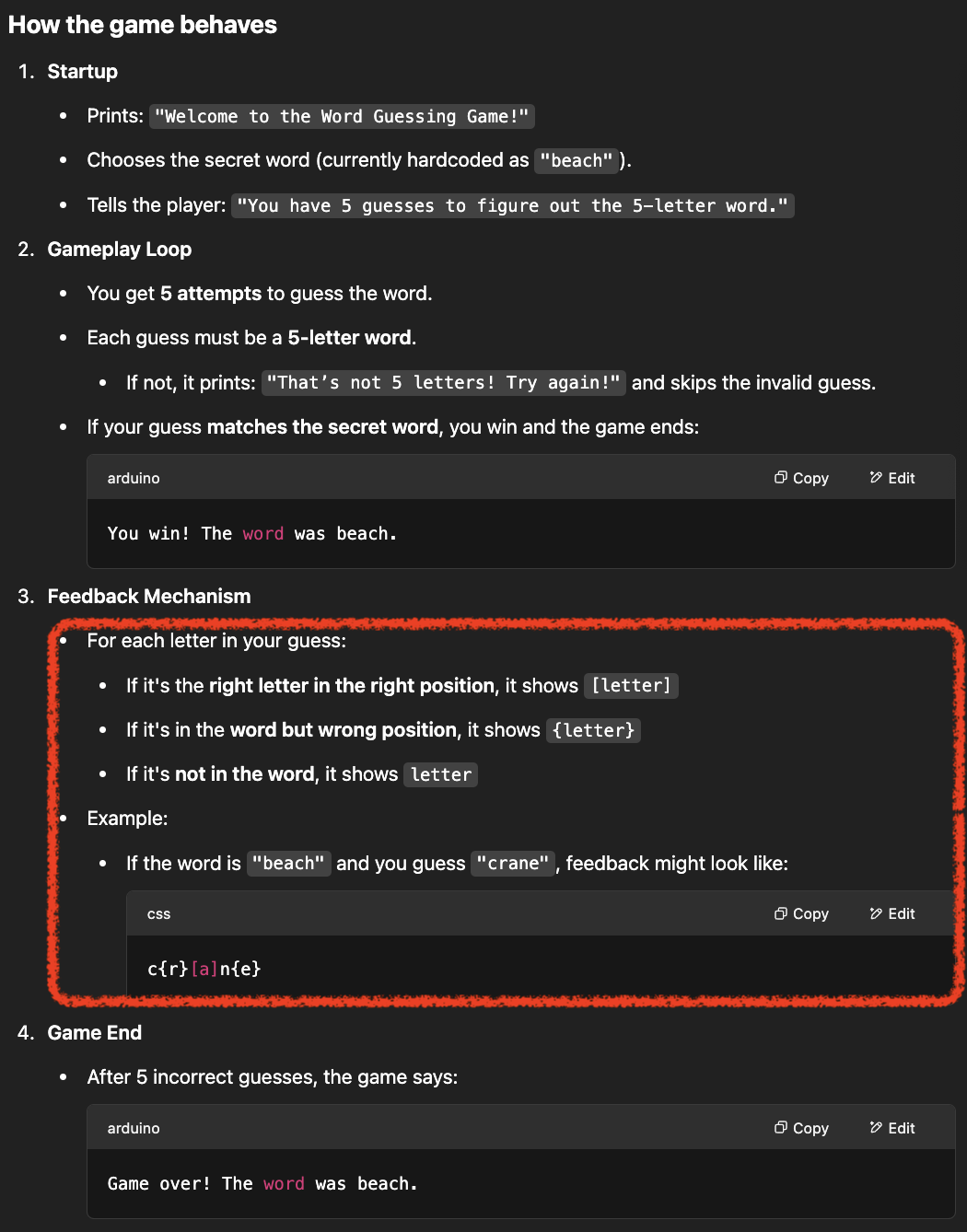}
  \caption{ChatGPT Hallucination for Wordle-like code} 
  \Description{}
  \label{fig:ASSETS_LLM_Hallucination}
\end{figure}

To our (unpleasant) surprise, ChatGPT hallucinated terribly, as shown in Figure \ref{fig:ASSETS_LLM_Hallucination}.

We are certain that future versions of LLMs will only get better and hallucinate less frequently.
But, BLV developers need better interaction support to help think about dynamic program behavior as well, without having to explicitly ask for it.
It is especially critical that LLMs support the ``stepping through'' behavior of debugging, which has previously been inaccessible to them and has possibly contributed to them developing a ``blind-spot'' or not learning an important skill at all.

\textbf{Observation 2: Disambiguating error source }

Language compilers and interpreters provide the line numbers where they suspect the error might be in a file.
This information seemed to be insufficient for two BLV participants who chose to read the feedback from ChatGPT and make changes in the file themselves.

The scenario is presented when analyzing the second Python program. 
The program uses a recursive approach to solving a puzzle.
Recursive functions, by definition, call themselves to solve sub-problems in a manner that helps solve the main problem.

\begin{figure}[!htp]
  \centering
  \includegraphics[width=0.50\linewidth]{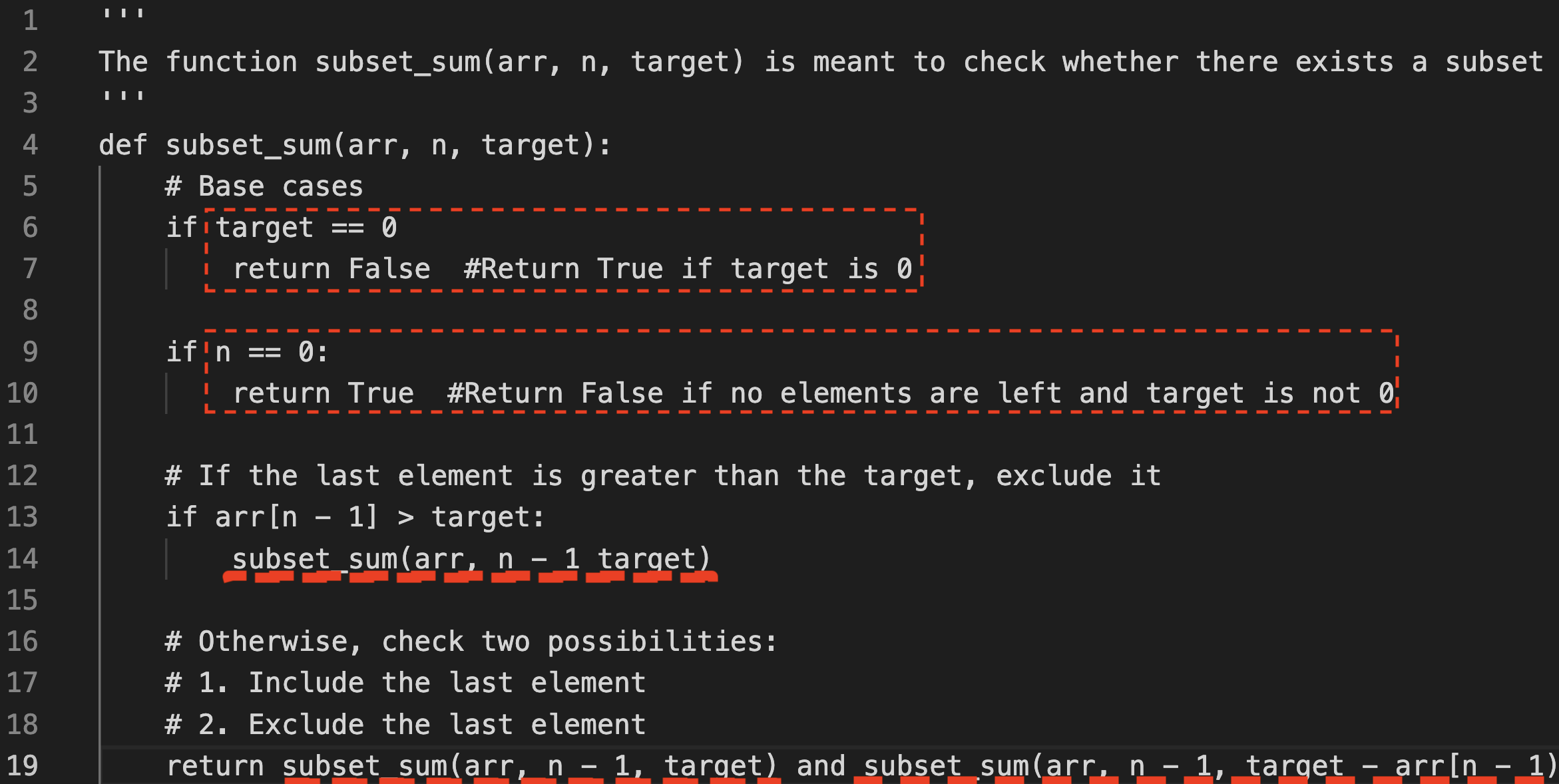}
  \caption{Areas in the Puzzle program with boxes highlighting the part of the code where ChatGPT hallucinated. The red marker highlight is the name of the recrusive function that ChatGPT did not disambiguate. }
  \Description{}
  \label{fig:ASSETS_Recur_LLM_Hallucination}
\end{figure}

Both ChatGPT and the program console said that there was a syntax error in a recursive function call in line 19 of the code (see Figure \ref{fig:ASSETS_Recur_LLM_Hallucination}).
Sighted users can instantaneously observe that there are two recursive function calls in the same line.
P4 and P7 did not know that the code is written in that manner and on hearing the code from left to right, stopped with the first recursive function call and expressed confusion because what they were hearing from the screen reader did not align with the error that ChatGPT/console highlighted. 
Eventually, P4 figured out that there was a second recursive function call in the same line. 
P7 noted that they, as a habit, ``\textit{never paused to hear line numbers because it is often unreliable.}’‘
They instead chose to find the errors by ``searching'' through code base for the erroneous code.
This programming practice made matters worse because the function was recursively called in three different places.
For a sighted user, it is easier to catch that.
P7 instead (incorrectly) made changes to another instance of the function call. 
This threw them off entirely and they were not able to recover from it (even with subsequent querying to ChatGPT).

Line numbers are shown to potentially help developers find the source of the error quickly.
But, clearly it is designed for developers who can visually skim code. 
We hypothesized that ChatGPT, or any ``smart'' system could do more than a standard compiler or interpreter.
In fact, what threw P4 even more was that their initial prompt of ``tell me what this code does'' resulted in ChatGPT listing the errors, but not listing any line numbers.
They had to query ChatGPT, subsequently, to obtain that information.
This, along with ChatGPT hallucinating (see below) caused P4 to stop and remark that this code is ``\textit{above my pay grade.}''

(Minor) Observations:
\begin{itemize}
 \item 1. \textbf{ChatGPT said it, must be right}: Figure \ref{fig:ASSETS_Recur_LLM_Hallucination} shows part of the code where the code does not match the corresponding comment. 
 It was the only time in the entire study that ChatGPT hallucinated and incorrectly suggested to P4 that the comment was ``misplaced.'' 
 In response, P4 deleted the correctly placed comment instead of matching the code to comment. 
 This action had a cascading effect as P4 quickly went down a rabbit hole that they should not have been in the first place. Providing the ability to simulate program behavior would have helped tremendously as compared to static program analysis.
 \item 2. \textbf{Having comments in code helps! right?}: Even though both code files presented to the participants included detailed comments, they chose to scroll past the comments without explicitly engaging with them, instead focusing on the code. 
 By skimming the comments, sighted users would quickly have all the information they need to answer both questions asked in the study without any need to ask ChatGPT to help with code comprehension.
 \item 3. \textbf{Sub-tasks autonomy}: Across both programming tasks, participants did not appear to use ChatGPT to support code navigation or skimming.
 \item 4. \textbf{Information overload}: Long unstructured traceback errors, program outputs or logs are extremely challenging for BLV users to parse. 
 If/When participants were not specific in their prompt construction, ChatGPT did not seem to know if they wanted information to solve the bug or to understand it. 
 In an effort to be helpful, it provided both assuming that the user \textbf{could} skim to the part of the response that they found pertinent. 
 Just like that, BLV needs are ignored. Without quick resolution, BLV developers are likely to adopt workarounds, an unnecessary skill, to prevent cognitive overload.
\end{itemize}

\section{Conclusion} \label{conclusion}
It is extremely unfortunate that we are seeing history repeat itself: new technology meant to improve performance in coding; but not designed accessibly from the ground-up so as to make the field more equitable.
Through review of most recent literature, we identified scenarios where LLMs would help BLV developers.
At the same time, we identified newer challenges for the BLV community as a result of working with LLMs.
Our exhaustive evaluation of LLM-integrated IDEs on programming tasks similarly showed the promise of LLMs as well as inconsistencies in its performance that would have a more significant impact on BLV developers.
From our user study, we highlighted several shortcomings in LLMs support as a part of a BLV developer's programming workflow.
Some of them, minor in scale, would naturally be solved as models improve.
But, others require serious thought and (re-)design effort from the companies and developers.
If we perform re-designs to improve and support interactions of BLV developers in their programming experience, those developers can have the same experience and opportunities as sighted individuals.

\begin{acks}
To the National Fed. of Blind for helping us recruit participants for the study and to each of the participants who shared their experiences with us.
\end{acks}


\end{document}